# Topological Magneto-optical Kerr Effect without Spin-orbit Coupling in Spin-compensated Antiferromagnet

Camron Farhang,[1,+] Weihang Lu,[1,+] Kai Du,[2] Yunpeng Gao,[3] Junjie Yang,[3] Sang-Wook Cheong,[2] and Jing Xia[1,*]

[1] Department of Physics and Astronomy, University of California, Irvine, Irvine, CA 92697, USA

[2] Keck Center for Quantum Magnetism and Department of Physics and Astronomy, Rutgers University, Piscataway, NJ 08854, USA.

[3] Department of Physics, New Jersey Institute of Technology, Newark, New Jersey 07102, USA

+These authors contributed equally.

*Correspondence: xia.jing@uci.edu

**Abstract**

The magneto-optical Kerr effect (MOKE), the differential reflection of oppositely circularly polarized light, has traditionally been associated with relativistic spin-orbit coupling (SOC), which links a particle's spin with its orbital motion. In ferromagnets, large MOKE signals arise from the combination of magnetization and SOC, while in certain coplanar antiferromagnets, SOC-induced Berry curvature enables MOKE despite zero net magnetization. Theoretically, large MOKE can also arise in a broader class of magnetic materials with compensated spins, without relying on SOC - for example, in systems exhibiting real-space scalar spin chirality. The experimental verification has remained elusive. Here, we demonstrate such a SOC- and magnetization-free MOKE in the noncoplanar antiferromagnet $Co_{1/3}TaS_2$. Using a Sagnac interferometer microscope, we image domains of scalar spin chirality and their reversal. Our findings establish experimentally a new mechanism for generating large MOKE signals and position chiral spin textures in compensated magnets as a compelling platform for ultrafast, stray-field-immune opto-spintronic applications.

## Introduction

The magneto-optical Faraday [1] and Kerr [2] effects, which reflect the fundamental interactions between circularly polarized light and magnetic materials, have been known for over a century. Among them, the magneto-optical Kerr effect (MOKE) has been extensively utilized for probing magnetic properties [3], visualizing magnetic domains [2], manipulating magnetic order [4], and enabling a variety of magneto-optical technologies [5], even down to the two-dimensional (2D) limit [6,7].

Spin-orbit coupling (SOC) [8,9], a relativistic interaction between a particle's spin and its orbital motion, has been deemed central in generating large MOKE signals. The presence of MOKE in ferromagnets has been attributed to the interplay between SOC and band exchange splitting (BES) in the band structure [10] induced by either the spontaneous magnetization or an external magnetic field. In non-collinear antiferromagnets with negligible net magnetization [11], large MOKE signals up to 300 μrad have recently been predicted [12] and observed [13], due to a non-zero net Berry curvature in the band structure induced by SOC. This dependence on SOC appears to be a general requirement: although finite MOKE can be observed in systems such as time-reversal symmetry-breaking superconductors [14–16] and orbital Hall materials [17] with negligible SOC, the magnitude of the effect in these cases is typically several orders of magnitude smaller, often in the nanoradian range. With the emergence of novel magnetic materials such as altermagnets [18,19], where SOC and net magnetization are both intrinsically weak or absent, it becomes a question of fundamental significance whether large MOKE signals can be achieved without relying on SOC or net magnetization. This is also of practical importance, as MOKE offers a practical readout for spintronic and opto-spintronic devices based on altermagnets and antiferromagnets of compensated spins, with minimal stray fields, field robustness, and fast switching [20–23].

Theoretically, spin-dependent band splitting, and thus MOKE, can also arise from real-space spin textures of compensated spins without referencing relativistic SOC [24–26]. For instance, non-relativistic exchange-driven band splitting occurs in itinerant electron systems, giving rise to phenomena like altermagnetism [18,19]. One of the simplest examples is a noncoplanar antiferromagnet, proposed to exhibit a novel topological light-matter interaction [24], inducing a topological MOKE signal due to finite scalar spin chirality without relying on SOC or BES, i.e. without a net magnetization. In its minimal configuration as illustrated in Fig.1a, three noncoplanar twisted spins $S_i$, $S_j$, and $S_k$ generate a fictitious U(1) gauge field $b_f \propto t_3 \chi_{ijk} \hat{n}$, where $t_3 = t_{ij} t_{jk} t_{ki}$ is the product of successive electron hopping integrals around the triangular loop $i \rightarrow j \rightarrow k \rightarrow i$, $\chi_{ijk} = S_i \cdot (S_j \times S_k)$ is the scalar spin chirality representing real-space "spin-winding", and n̂ is the unit vector normal to the face of the triangle. This fictitious field $b_f$ originates from the orbital motion of electrons: as an electron hops around the triangle, it acquires a geometric Berry phase [27] corresponding to the solid angle subtended by the three spins. This topologically nontrivial Berry phase results in a phase difference $\Delta\varphi$ between left- and right-circularly polarized (LCP and RCP) light upon reflection, manifesting as a Kerr angle $\theta_K = \Delta\varphi/2$. Notably, this coupling between spin and orbital degrees of freedom (spin-dependent band splitting) arises from Kondo exchange coupling between conduction electrons and a noncoplanar arrangement of localized magnetic moments, without requiring relativistic SOC [25].

Despite the promising potential of this SOC-free mechanism for MOKE in compensated magnetic systems, experimental verification has remained elusive. Proposed candidate systems such as γ-$Fe_xMn_{1-x}$ and γ-$Na_xCoO_2$ [24] have not exhibited the expected signatures. This absence is partly due to the limited exploration of non-coplanar antiferromagnetic orders and partly because the fictitious magnetic field arising from scalar spin chirality often cancels out due to symmetry, typically requiring external strain to stabilize a net effect [24]. Lastly, the choice of the optical wavelength is non-trivial as the theoretically calculated Kerr signal oscillates strongly with the photon energy, crossing zero frequently [24].

In this study, we report the first experimental realization of a large spontaneous MOKE signal of $250\ \mu rad$ at the technologically relevant telecommunication wavelength of $1550\ nm$ in the triangular lattice compound $Co_{1/3}TaS_2$ [28–30], despite its negligible net magnetization of $0.01\ \mu_B/Co$. This result confirms the viability of the SOC-free mechanism for producing technologically relevant MOKE signals. Using MOKE imaging, we visualize scalar spin chirality domains and their switching behavior under applied magnetic fields.

## Results

$Co_{1/3}TaS_2$ belongs to a broader family of magnetic element-intercalated van der Waals transition metal dichalcogenides, which exhibit a wide range of magnetic states depending on the intercalants [31–36]. Its crystal structure consists of chiral alternating layers of cobalt, which carry localized magnetic moments, and metallic tantalum disulfide, which hosts itinerant electrons. While neutron scattering studies in the 1980s reported [33] a magnetic modulation vector of $q_m = (1/3, 1/3, 0)$, leading to interpretations involving a "toroidal" magnetic order [28], recent experiments [30,29] on $Co_{1/3}TaS_2$ and $Co_{1/3}NbS_2$ have instead identified $q_m = (1/2, 0, 0)$. The latter corresponds to a non-coplanar "triple-Q" order [30,29], simultaneously breaking time-reversal and inversion symmetry. Consequently they exhibit anomalous Hall effect [37,30,29] and Nernst effect [38] arising from an uncompensated fictitious magnetic field, even though their net magnetic moments are vanishingly small. A clear correlation between the degree of spin non-coplanarity and the amplitude of spontaneous Hall signal was confirmed by temperature dependence of polarized neutron scattering [30]. As illustrated in Fig.1c, the magnetic unit cell in this triple-Q state includes eight triangular plaquettes that generate a strong fictitious field oriented along the crystallographic c-axis. Importantly, this magnetic order in an ideal triple-Q state [29] does not involve any net spin magnetization; rather, the fictitious field couples to the orbital motion of itinerant electrons, producing a small orbital magnetic moment of approximately $0.01\ \mu_B$ per $Co^{2+}$ ion [29,30]. Note that the real material $Co_{1/3}TaS_2$ may deviate from this ideal triple-Q state, and the observed zero-field magnetization could include some spin contribution.

This spin configuration has non-zero scalar spin chirality and is not associated with any spin rotation symmetry, so non-relativistic spin splitting can occur along any direction. In addition, this spin configuration can have a net magnetization along $c$-axis when SOC is included, so it may be considered as an "M-type" altermagnet with magnetic point group of 32' [39].

Upon warming, this triple-Q order transitions at the Néel temperature $T_{N2}$ into a single-Q stripe order as illustrated in Fig.2b with zero scalar spin chirality, and subsequently into a paramagnetic state above $T_{N1}$ [29]. The $Co_{1/3}TaS_2$ crystals used in our experiments were synthesized via chemical vapor transport (Methods), as shown in the inset of Fig.1d. Magnetic susceptibility measurements under field-cooled (FC) and zero-field-cooled (ZFC) conditions (Fig.1b) indicate transition temperatures of $T_{N1}{\sim}33\ K$ and $T_{N2}{\sim}16\ K$.

MOKE detection and imaging are performed using a zero-loop Sagnac interferometer microscope [6,14,40,41] in the polar geometry (Fig.1e) operating at the most popular telecommunication wavelength of $1550\ nm$ ($0.80\ eV$ photon energy). This technique offers exceptional sensitivity, routinely achieving $0.01\ \mu rad$ resolution. This is made possible by its exclusive detection of microscopic time-reversal symmetry breaking (TRSB) while strongly rejecting non-TRSB effects, such as optical birefringence, with a suppression level of $10^{-6}$ (Supplementary Information). This distinction is particularly important in $Co_{1/3}TaS_2$, where some of the authors have recently observed a birefringent polarization rotation of $600\ \mu rad\ sin(2\alpha)$ due to a nonvolatile nematic order, with $\alpha$ being the incident polarization angle [42]. The selective sensitivity to TRSB is achieved by using a single-mode optical fiber as both the source and detector for counter-propagating, time-reversed light beams [43]. According to Onsager's relations [44], this configuration guarantees zero signal in the absence of TRSB.

We now present the results of our MOKE measurements. Fig.1d shows the spontaneous MOKE signal $\theta_K$ obtained during zero-field warm (ZFW) after $B = 0.3\ T$ field cool (FC), measured at a single point with $2\ \mu m$ optical beam size. Remarkably, we observe a giant $\theta_K$ of $250\ \mu rad$, comparable to the $300\ \mu rad$ signal reported in the coplanar non-collinear antiferromagnet $Mn_3Sn$ [13], where the large MOKE arises from relativistic SOC-induced Berry curvature [12]. It is important to note that both $Co_{1/3}TaS_2$ and $Mn_3Sn$ exhibit vanishingly small net magnetic moments, with values of $\sim 0.01\ \mu_B/Co$ [29,30] and $\sim 0.005\ \mu_B/Mn$ [45], respectively. This experimental result thus firmly establishes that the SOC-free mechanism driven by the scalar spin chirality can produce MOKE signals on par with those generated by SOC-induced Berry curvature. The pronounced temperature dependence of MOKE in $Co_{1/3}TaS_2$ stands in sharp contrast to the temperature-invariant behavior we observed in the coplanar non-collinear antiferromagnet $Mn_3NiN$ [46], underscoring their distinct origins. When compared to theoretical predictions, the calculated $\theta_K$ in the proposed triple-Q state of γ-$Fe_xMn_{1-x}$ is even larger, reaching $\pm 10\ mrad$ depending on the photon wavelength [24]. These observations suggest that the SOC-free mechanism holds substantial promise for approaching the largest reported MOKE of $\sim 10\ mrad$ in ferromagnets CoPt [47].

In contrast to the large $\theta_K$ in the triple-Q state, no measurable MOKE signal was detected during ZFW through $T_{N2}$ into the single-Q phase, or upon further warming above $T_{N1}$ into the paramagnetic phase (Fig.1d). This is consistent with theoretical predictions: both the single-Q and paramagnetic phases lack scalar spin chirality and thus are not expected to support a finite MOKE [24].

We observed negligible spontaneous $\theta_K$ after zero-field cool (ZFC) (Supplementary Fig. 7b), indicating the formation of oppositely polarized domains smaller than the optical beam size in the absence of a training field. This also confirms that the zero-field MOKE signal does not originate from uncompensated spins at AFM domain walls. If that were the case, ZFC, which produces a higher density of domain walls, would yield a larger MOKE signal than field cooling (FC), contrary to our observations. The negligible magnetic contribution from AFM domain walls is also supported by MFM measurements after ZFC (Supplementary Fig. 8), which show no detectable fringing fields from domain walls.

Fig.2a summarizes the MOKE hysteresis loops across different magnetic phases of $Co_{1/3}TaS_2$. At $50\ K$, in the paramagnetic phase, the Kerr signal $\theta_K$ exhibits a small linear response to the applied magnetic field $B$, consistent with partial alignment of magnetic moments by the external field. At $20\ K$, in the single-Q phase as illustrated in Fig. 2b, the MOKE signal develops a shallow S-shaped curve but remains zero at zero field, consistent with the absence of scalar spin chirality $\chi_{ijk}$. Upon cooling to $15\ K$, as the system enters the non-coplanar triple-Q phase, a small hysteresis loop emerges between $\pm 0.4\ T$ (cyan curve in Fig. 2a), signaling the onset of a finite $\chi_{ijk}$ and the associated fictitious field $b_f$. Notably, across $\sim 5\ T$ magnetic field, there is a smooth change in the Kerr signal, corresponding to a metamagnetic transition into a different magnetic configuration [29,30], referred to here as the triple-Q' state. Further cooling to $10\ K$, $5\ K$, and $2\ K$ leads to increasingly wide and pronounced hysteresis loops, indicating higher coercive fields required to reverse the chirality domains and stronger fictitious fields at zero applied field. The jump in MOKE signal across $5\ T$ field also becomes more abrupt at lower temperatures, signifying a sharper metamagnetic transition between the triple-Q and triple-Q$'$ phases.

The most striking feature of the Kerr hysteresis in the triple-Q phase (Fig.2a) is its field independence: once the sign of $\chi_{ijk}$ is established by the external field, the Kerr signal remains nearly constant between $0\ T$ and $4\ T$ fields. A similar plateau is observed in the triple-Q' phase above $6\ T$. For clarity, the $2\ K$ MOKE hysteresis is replotted in Fig. 2d, revealing a stark contrast to the corresponding $2\ K$ magnetization $M_z$ (Fig. 2c). $M_z$ begins at $\sim 0.01\ \mu_B$ per Co at zero field and increases linearly at a rate of $dM_z/dB \sim 0.02\ \mu_B/T$, except for a discrete jump of $\sim 0.07\ \mu_B$ at $5\ T$ due to the metamagnetic transition.

We estimate that the magnetization contribution to the MOKE signal is negligible, less than 1% (see Supplementary Information). This contrasting field dependence between MOKE and $M_z$ confirms their decoupling: while MOKE arises from the fictitious field $b_f$ generated by scalar spin chirality, the magnetization includes both spin and orbital contributions. Specifically, the zero-field $M_z$ and its linear slope $dM_z/dB$ are due to the orbital moment of itinerant electrons, induced respectively by the fictitious field $b_f$ and the external magnetic field $B$. These orbital contributions are not directly coupled to the Kerr signal. In contrast, the metamagnetic transition, which involves a spin reconfiguration, produces simultaneous step-like changes in both $M_z$ and $\theta_K$. Importantly the optical reflectivity remained constant to within 0.5% during the entire measurement (Fig. 2g), confirming that the metamagnetic transition represents a change of the spin configuration and has minimum impacts on the electronic structure.

Since the nature of the metamagnetic transition has not been resolved by neutron scattering experiments yet [29,30], we adopt a simplified working assumption: the transition reflects a sudden change in the angle $\theta$ between spins $S_1$ and $S_4$, as illustrated in the inset of Fig. 2d. Under this assumption, the net spin moment is given by $M_{\mathrm{spin}} = \frac{1}{2}(-1 - 3\cos(\theta))\,\mu_B$, assuming a *g-factor* of -2. And it can be estimated experimentally from magnetization as $M_{\mathrm{spin}} \sim M_z - 0.02\,\mu_B \cdot B$, accounting for the orbital contribution. The scalar spin chirality $\chi_{ijk}$ is assumed to scale with the bottom triangular plaquette $\chi_{ijk} \propto \chi_{123} \propto -\frac{3}{2}\sqrt{3}\cos(\theta)\sin(\theta)^2$. Within this simplistic model, we estimate $\chi_{ijk} \sim -\frac{2(1+2M_{spin})(-2+M_{spin}+{M_{spin}}^2)}{3\sqrt{3}}$, as plotted in Fig. 2f. Comparing this estimated $\chi_{ijk}$ and the measured $\theta_K$ (Fig. 2d), the model qualitatively captures the key features of the magnetic-field dependence of MOKE. The quantitative discrepancy across the transition across $5\,T$ suggests that a more accurate model is required for this metamagnetic transition. This will depend on resolving the spin structure of the triple-Q' phase in future neutron scattering experiments, beyond the scope of the present work. Lastly, we note the DC Hall effect $\sigma_{xy}$ (Fig. 2e) shows an opposite change across the metamagnetic transition, suggesting a rather complex frequency dependence related to the corresponding change of spin texture across the metamagnetic transition.

Spin chirality can be imaged by MOKE microscopy that provides a powerful, non-contact method for mapping spin chirality domains and their associated dynamics. One immediate application is the assessment of chemical inhomogeneity. Recent neutron scattering and transport studies [48] have reported that even slight variations in cobalt composition can lead to significant changes in the physical properties of $Co_xTaS_2$. It is therefore critical to determine the extent of chemical inhomogeneity, and its impact on the MOKE signal within a single crystal. We found that higher Co content leads to lower red-light reflectivity, an effect which we expect to persist at 1550 $nm$ in the Sagnac measurements. (Supplementary Fig. 9 c, d).

In Fig.3, we present a combined study of optical reflectivity imaging, MOKE imaging, and single location MOKE hysteresis, all performed at 2 $K$. The optical reflectivity map (Fig. 3b) is largely uniform with $0.42 < R < 0.44$, which we use as a convenient proxy for cobalt composition as it is difficult to perform energy dispersive x-ray spectroscopy (EDX) chemical mapping in the same region. MOKE hysteresis loops measured at a few representative locations with varying reflectivity values are shown in Fig. 3a. All locations exhibit nearly identical coercive fields of $\sim 0.8\ T$, with the primary variation being in the zero-field Kerr signal (200, 202, 160, and 80 $\mu rad$, respectively). To further examine the magnetic inhomogeneity, we performed MOKE imaging at three key points along the hysteresis loop: at zero field (Fig. 3d), and near the switching fields at $-1\ T$ (Fig. 3c) and $+0.8\ T$ (Fig. 3e). Near the switching fields, large variations in both positive and negative Kerr signals are observed (Fig. 3c, e), consistent with the coexistence and reversal of spin chirality domains across the sample. Importantly, the spatial patterns seen in Figs. 3c and 3e do not resemble those in the reflectivity map (Fig. 3b), suggesting that the coercive field and domain switching behavior are not strongly correlated with cobalt composition. Instead, they are likely influenced by local strain, defects, or other extrinsic structural factors.

In stark contrast, a clear correlation emerges between the optical reflectivity (Fig. 3b) and the zero-field MOKE image (Fig. 3d), where the chirality has been uniformly trained to a negative value. This observation implies that cobalt composition primarily affects the magnitude of the MOKE signal in the zero-field state, rather than the switching dynamics. To further test this hypothesis, we investigated a rare inhomogeneous region (see Supplementary Fig. 1) where reflectivity varies significantly from 0.40 to 0.45. In this region, the spontaneous MOKE signal ranges from $-40\ \mu rad$ to $-180\ \mu rad$. A plot of MOKE vs. reflectivity in Supplementary Fig. 1 reveals an empirical linear relationship within this limited reflectivity range, supporting the idea that local cobalt concentration modulates the amplitude of the Kerr signal.

Having established that most of the sample is magnetically uniform, we now demonstrate the utility of MOKE microscopy in visualizing scalar spin chirality domain reversal under an external magnetic field. These measurements were conducted in magnetically homogeneous regions at an elevated temperature of 10 $K$, where domain reversal occurs over a broader field range. The results are shown in Fig.4 and Supplementary Fig. 2.

A sequence of MOKE images from a uniform region is presented in Figs. 4b-i, with the corresponding local hysteresis loop taken at a single location shown in Fig. 4a. At $B = -9\ T$ (Fig. 4b), the MOKE signal is uniformly negative ($\theta_K \sim -200\ \mu rad$, shown in blue), except for a few isolated spots that are likely regions of differing Co composition with near-zero Kerr signals (green). At $B = 0$ (Fig. 4c), patches with $\theta_K \sim 0$ begin to emerge. These are interpreted as sub-wavelength domains with mixed positive and negative chiralities, whose contributions average to near zero within the $\sim 2\ \mu m$ optical spot size. As the magnetic field increases (Figs. 4d-f), these green regions grow and are joined by new areas exhibiting positive chirality (orange to red). The positive domains progressively expand through domain wall motion, eventually dominating the lower half (Fig. 4g) and finally the entire imaged region (Fig. 4h-i), culminating in a uniform positive Kerr signal of $\theta_K \sim 200\ \mu rad$.

Notably, even at a low field of $0.1\ T$ (Fig. 4d), isolated positive domains are already present. This indicates that the coercive field for domain reversal is not fundamentally dictated by the intrinsic non-coplanar magnetic phase but is instead governed by extrinsic factors such as local strain, defects, or pinning centers. This insight suggests that by engineering these extrinsic factors, it may be possible to tailor coercive fields in $Co_{1/3}TaS_2$ and related materials, either minimizing them for low-power logic switching applications or maximizing them for robust spin memory devices.

Further evidence of this chirality domain reversal dynamics is provided in Supplementary Fig. 2, which presents MOKE images of another region taken at $1\ T$ intervals across the full field sweep. As in the main data, domain reversal near $\pm 1\ T$ occurs via domain wall motion, with positive and negative chirality domains coexisting during the transition.

Interestingly, Supplementary Fig. 2 also captures the distinct dynamics of the metamagnetic transition, which occurs gradually between $B = 4\ T$ and $6\ T$. Across the $3\ T \rightarrow 9\ T \rightarrow 3\ T$ cycle, the MOKE signal changes smoothly and uniformly across the field of view, with no evidence of domain wall motion. This behavior resembles a coherent rotation process in ferromagnetic hysteresis, and implies that the critical field for the metamagnetic transition is determined by the intrinsic spin structure transformation. This conclusion is supported by the consistent metamagnetic transition field observed across different samples and temperatures [30,29]. Finally, Supplementary Fig. 3 shows a zero-field MOKE image taken at $20\ K$ in the single-Q phase. The MOKE signal is uniformly zero across the imaged region, consistent with the expectation that scalar spin chirality, hence MOKE, is absent in the single-Q phase.

## Discussion

The observed topological MOKE in $Co_{1/3}TaS_2$ represents a fundamentally new form of light-matter interaction arising from the chirality of real-space spin textures composed of compensated spin moments, an effect predicted by theory [24] but not previously observed. Unlike conventional MOKE mechanisms that depend on relativistic spin-orbit coupling of individual spins or a net spin moment [10,12], this SOC-free mechanism is rooted in the collective topology of the real-space spin texture with compensated spins. It is therefore expected to apply to a much broader class of magnetic materials with "winding" real-space spin configurations. It is noted that although SOC is not required for the topological MOKE discussed here, it can slightly modify its strength, as shown in theoretical studies of topological Hall conductivity in the DC limit [49].

Supplementary Fig.4 compares the topological MOKE-over-magnetization ratio ($\theta_K/M$) in $Co_{1/3}TaS_2$ with other magnetic systems. In ferromagnets (Ni [50], Fe [50], and CoPt [47]), spontaneous $\theta_K$ scales with remanent magnetization due to SOC, following a $\sim 0.03\ rad/\mu_B$ ratio. In contrast, Noncoplanar antiferromagnet $Co_{1/3}TaS_2$ and coplanar antiferromagnet $Mn_3Sn$ [13] exhibit much larger $\theta_K/M$ ratios due to negligible net magnetization ($0.005 - 0.01\ \mu_B$). It is important to note that while MOKE in $Mn_3Sn$ is driven by SOC-induced Berry curvature, MOKE in $Co_{1/3}TaS_2$ arises from real-space scalar spin chirality. The skyrmion lattice phase in $Gd_2PdSi_2$ [51] has a topological contribution to MOKE from swirling spins, though limited to a narrow field range and a large net moment of $4\ \mu_B$, resulting in a small ratio $\sim 7 \times 10^{-5}\ rad/\mu_B$. Similarly, small ratios are found in skyrmion lattices in $CrVI_6$ [52] ($\theta_K \sim 2 \times 10^{-4}\ rad$, $M \sim 3\ \mu_B/Cr$) and graphene/$Fe_3GeTe_2$/graphene heterostructure [53] ($RMCD \sim 1 \times 10^{-3}\ rad$, $M \sim 2.5\ \mu_B/Fe$) primarily due to their large ferromagnetic moments.

Using the noncoplanar spin system in $Co_{1/3}TaS_2$ [28,30,29] as a model system, we demonstrated this mechanism's efficacy and employed MOKE microscopy to image both chiral domain reversal and the metamagnetic transition. The observed giant spontaneous MOKE at telecommunication wavelengths is readily detectable using standard experimental setups. Moreover, theoretical work predicts that at terahertz frequencies, this system can host a quantum topological Kerr effect with a quantized Kerr rotation of $\pi/2$ [24], surpassing all known magneto-optical materials in magnitude.

These findings have direct implications for advancing antiferromagnet and antiferromagnet-based spintronics and opto-spintronics [20–23]. Devices based on this SOC-free mechanism would inherently be immune to stray magnetic fields and capable of ultrafast switching, overcoming key limitations of traditional ferromagnetic technologies. MOKE is well suited for ultrafast, local detection of spin-chirality domain switching driven by chiral spin-orbit torques from electric current or circularly polarized light. More broadly, this mechanism relaxes the design constraints imposed by SOC-based magneto-optical materials, offering greater flexibility in engineering materials with tailored magneto-optical responses. We envision that SOC-free magneto-optical materials, especially the van der Waals layers, can be realized through diverse fabrication techniques, including nanofabrication, molecular beam epitaxy, layer stacking, and self-assembly. This opens a wide and largely unexplored landscape for next-generation magneto-optical materials and devices.

**Methods:**

**Crystal Growth:**

$Co_{1/3}TaS_2$ single crystals were grown by chemical vapor transport method. The powder samples of $Co_{1/3}TaS_2$ were synthesized first using the solid-state reaction method. High-purity powders of Cr (99.97%), Ta (99.9%), and S (99.999%) were mixed in stoichiometric ratios. The mixed powders were ground, pelletized, and sealed in quartz tubes. The pellets of powder were then sintered at 750 °C for 48 hours at a heating rate of ~15 °C/h, with intermediate grinding. For the single-crystal growth of $Co_{1/3}TaS_2$, the resulting black powder samples were sealed in evacuated quartz tubes together with iodine (I2) as a transport agent. The tubes were placed in a two-zone furnace for 10 days, where the hot and cold ends were maintained at 1000 °C and 900 °C, respectively. The resulting crystals are hexagonal plate-like.

**Magnetization measurements:**

The magnetic susceptibility versus temperature (χ-T) curves were measured using a Cryogenic-Limited Cryogen Free Measurement System (CFMS) with a vibrating sample magnetometer (VSM) option. A magnetic field of 0.1 T was applied along the c-axis of the $Co_{1/3}TaS_2$ single crystal during the measurements of χ-T curves. The magnetic moment versus magnetic field (M-H) curves were also measured using the VSM option of our CFMS. The crystal was first cooled to base temperature in a zero magnetic field, then the M-H curve was measured by ramping up the magnetic field.

**Hall measurements:**

The Hall effects of the sample were characterized by a Keithley 2182 Nanovoltmeter and Keithley 6221 Current Source, and the sample temperature and magnetic field were controlled by the CFMS system. During the measurement, an electric current of 1 mA was applied along the ab plane of the crystal, and the magnetic field was applied to the c-axis of the crystal.

**Sagnac MOKE measurements:**

The MOKE measurements are performed using a zero-loop fiber-optic Sagnac interferometer [43] operating at 1550 $nm$ wavelength. For MOKE imaging we utilize a scanning Sagnac microscope with 2 $\mu m$ lateral spatial resolution installed inside a cryostat with 1.8 $K$ base temperature and 9 $T$ magnetic field capability. The operation of the interferometer is described in the Supplementary Information.

**Low-temperature magnetic force microscopy (MFM):**

After zero-field cooling from room temperature, a freshly-cleaved $Co_{1/3}TaS_2$ crystal was scanned at 5 $K$ using a temperature-variable AFM system (Attocube) in a dual pass mode (lift height ~ 35 $nm$) with commercial Co/Cr-coated magnetic tips.

**Data availability:**

Source data are provided with this paper. They have been deposited in a figshare repository with https://doi.org/10.6084/m9.figshare.29583293.

## Acknowledgements

We thank J.G. Zheng for Energy-dispersive X-ray spectroscopy (EDX) at UC Irvine. This project was supported by NSF award DMR-2419425 and the Gordon and Betty Moore Foundation EPiQS Initiative, Grant # GBMF10276 awarded to J.X.. The work at Rutgers University was supported by the DOE under Grant No. DOE: DE-FG02-07ER46382 awarded to S.W.C.. J.Y. acknowledges support by DOE under Grant No. DOE: DE-SC0021188 awarded to J.Y.. The authors acknowledge the use of facilities and instrumentation at the UC Irvine Materials Research Institute (IMRI), which is supported in part by the National Science Foundation through the UC Irvine Materials Research Science and Engineering Center (DMR-2011967).

## Author Contributions

J.X. conceived and supervised the project. C.F., W.L., and J.X. carried out the optical measurements. K.D., Y.G., J.Y., and S.W.C. grew the crystals and carried out transport, magnetization, and magnetic force microscopy measurements. J.X. drafted the paper with the input from all authors. All authors contributed to the discussion of the manuscript.

## Competing interests

The authors declare no competing interest.

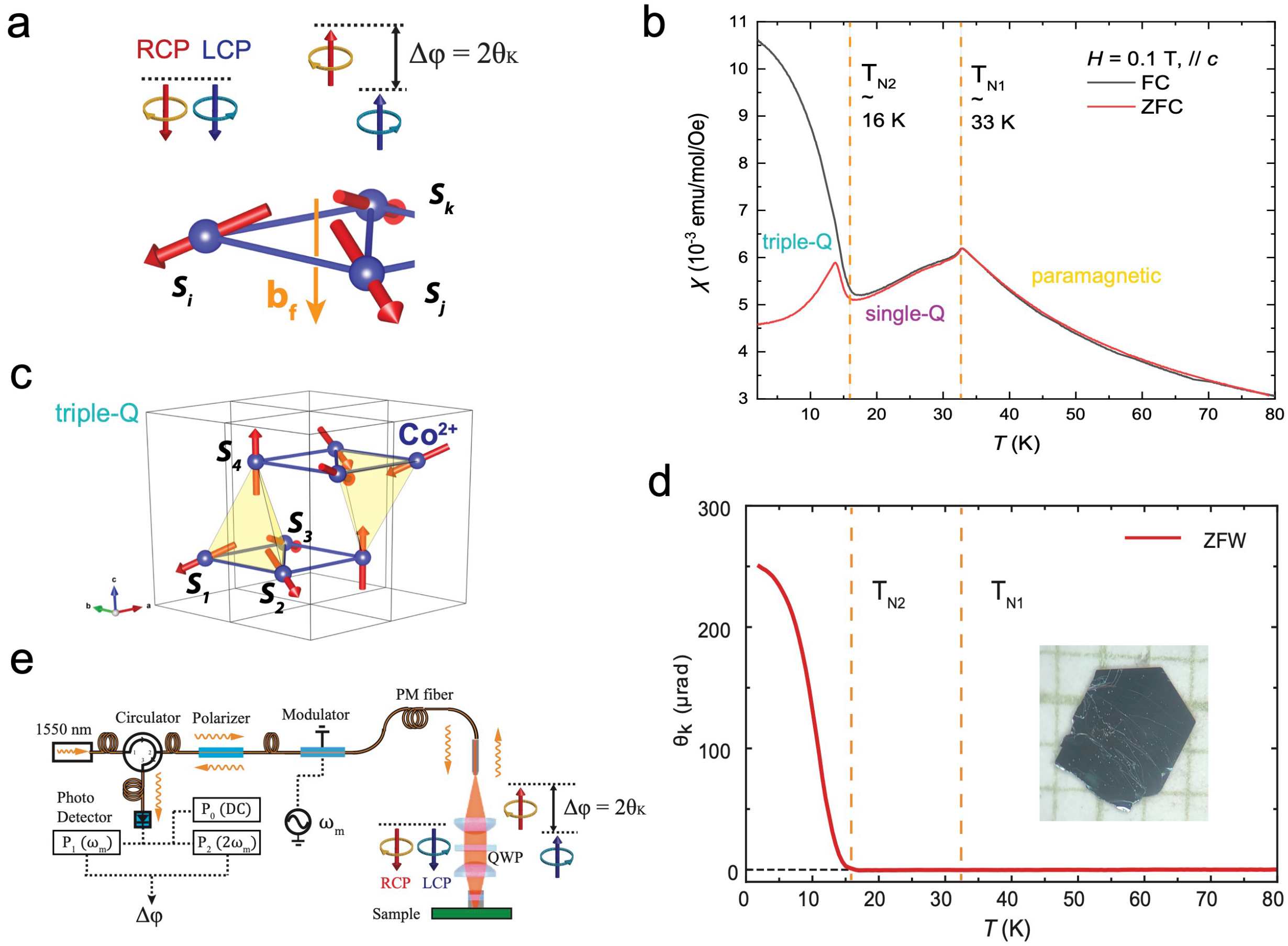

**Figure 1. Large MOKE in non-coplanar triple-Q state of $Co_{1/3}Ta_1S_2$. (a)** MOKE due to a new topological light-matter interactions in a minimal chiral magnet comprising three neighboring noncoplanar spins. $b_f$ is the fictitious magnetic field generated by the scalar spin chirality $\chi_{ijk} = S_i \cdot (S_j \times S_k)$. MOKE $\theta_K$ is a result of the phase difference $\Delta\varphi = 2\theta_K$ between reflected left and right-circularly polarized (LCP and RCP) light. **(b)** Magnetization $M_z$ measured after field cool (FC) and zero-field cool (ZFC). **(c)** The tetrahedral triple-Q state of $Co_{1/3}Ta_1S_2$. **(d)** MOKE $\theta_K$ measured during ZFW after $0.3\,T$ FC. Inset is a sample photo on $1\,mm$ grid paper. **(e)** Schematics of a zero-area-loop Sagnac interferometer microscope for polar MOKE measurements.

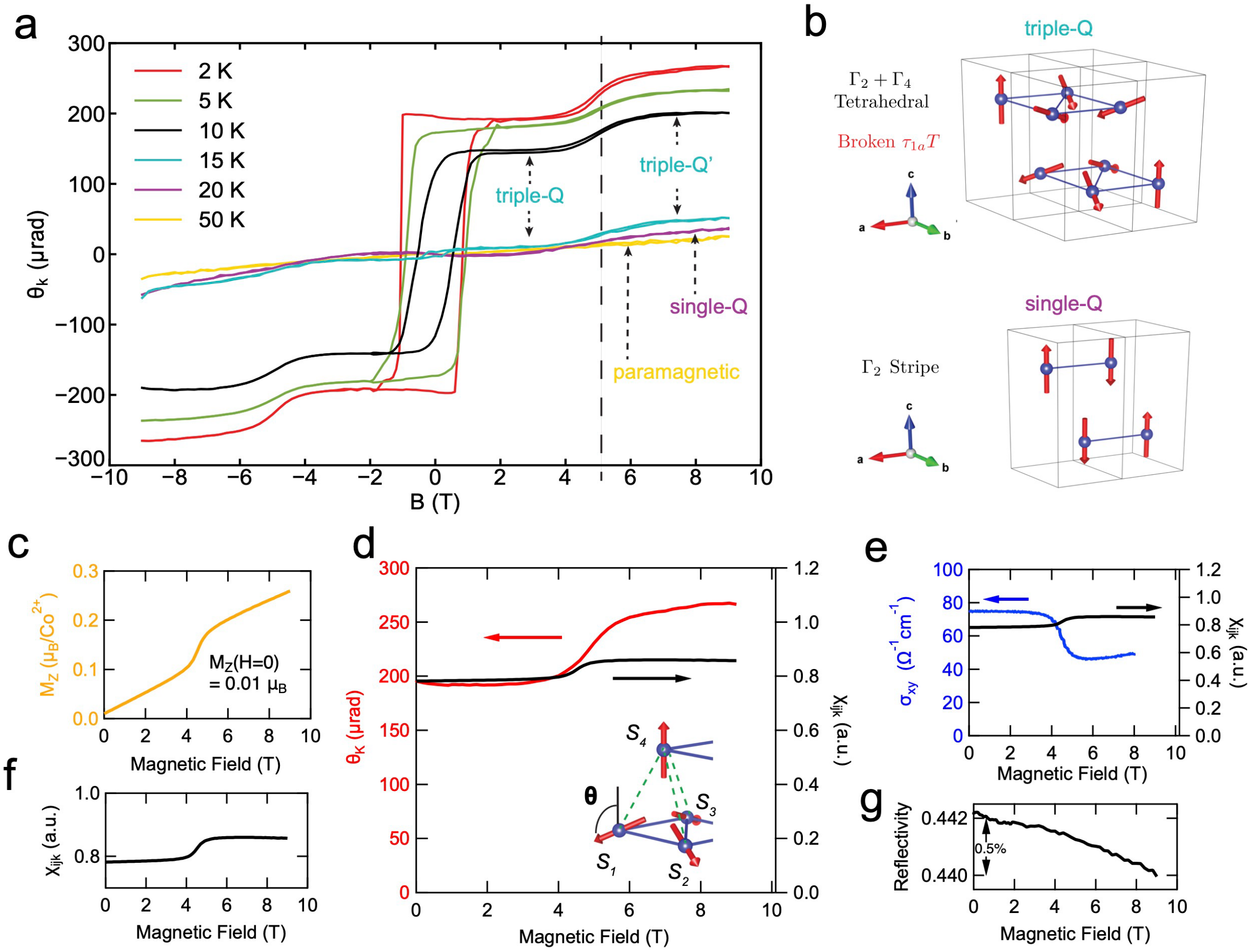

**Figure 2. MOKE hysteresis. (a)** MOKE hysteresis taken at spot 1 at various temperatures. **(b)** sketches of triple-Q and single-Q phases. **(c)** Magnetization $M_z$ at 2 $K$. **(d)** MOKE at 2 $K$ plotted with the estimated spin chirality $\chi_{ijk}$. **(e)** Hall conductivity $\sigma_{xy}$ at 2 $K$ plotted with $\chi_{ijk}$. **(f)** Spin chirality $\chi_{ijk}$ estimated from magnetization. **(g)** Reflectivity at 2 $K$.

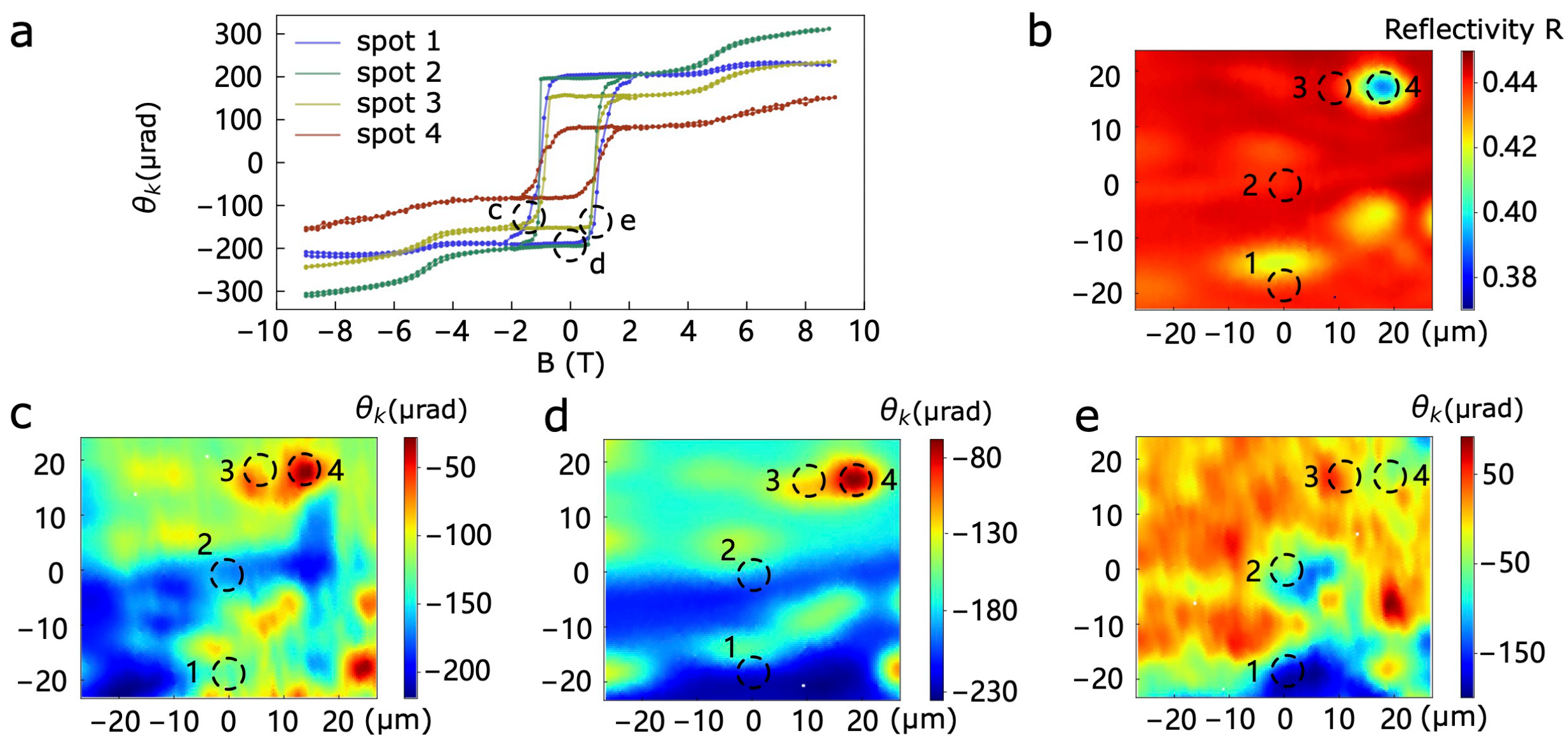

**Figure 3. Scalar spin chirality domains imaged at 2$K$. (a)** MOKE hysteresis measured at spots 1-4 at $T = 2\ K$. **(b)** Optical reflectivity image at $T = 2\ K$. **(c), (d), (e)** MOKE $\theta_K$ images taken at B = -1 T, 0 T, and 0.8 T respectively during the hysteresis loop at $T = 2\ K$, showing the formation of scalar spin chirality $\chi_{ijk}$ domains.

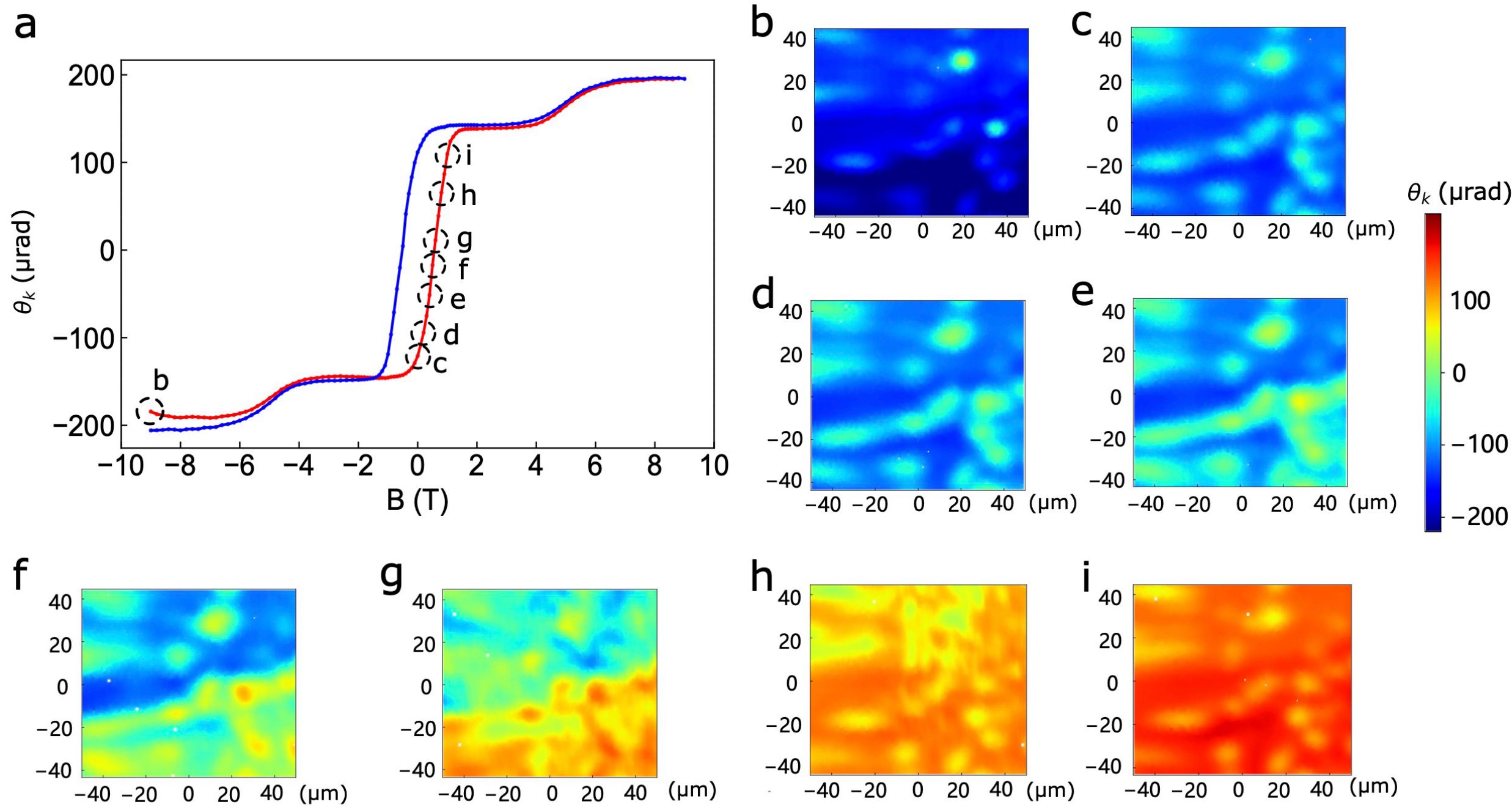

**Figure 4. Chirality domain reversal imaged at 10 *K*. (a)** MOKE hysteresis measured at a single spot of the imaged region at $T = 2\ K$. **(b) - (i)** MOKE images of chirality domain reversal via domain wall motion at $T = 10\ K$.

Supplementary Information for

# Topological Magneto-optical Kerr Effect without Spin-orbit Coupling in Spin-compensated Antiferromagnet

Camron Farhang,[1,+] Weihang Lu,[1,+] Kai Du,[2] Yunpeng Gao,[3] Junjie Yang,[3] Sang-Wook Cheong,[2] and Jing Xia[1,*]

[1] Department of Physics and Astronomy, University of California, Irvine, Irvine, CA 92697, USA

[2] Keck Center for Quantum Magnetism and Department of Physics and Astronomy, Rutgers University, Piscataway, NJ 08854, USA.

[3] Department of Physics, New Jersey Institute of Technology, Newark, New Jersey 07102, USA

+These authors contributed equally.

*Correspondence: xia.jing@uci.edu

**Contents:**

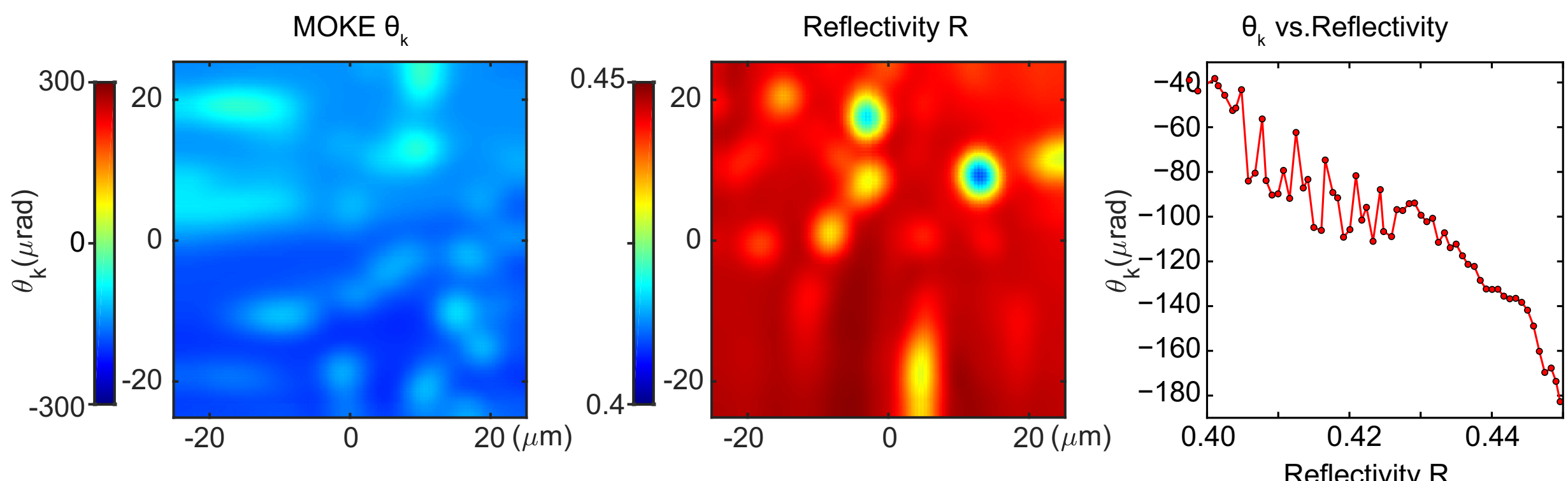

**Supplementary Figure 1. Extended data: Imaging the correlation between Co composition and magnetism at 10 $K$.** Spontaneous MOKE image (left), reflectivity image (middle), and their empirical correlation (right).

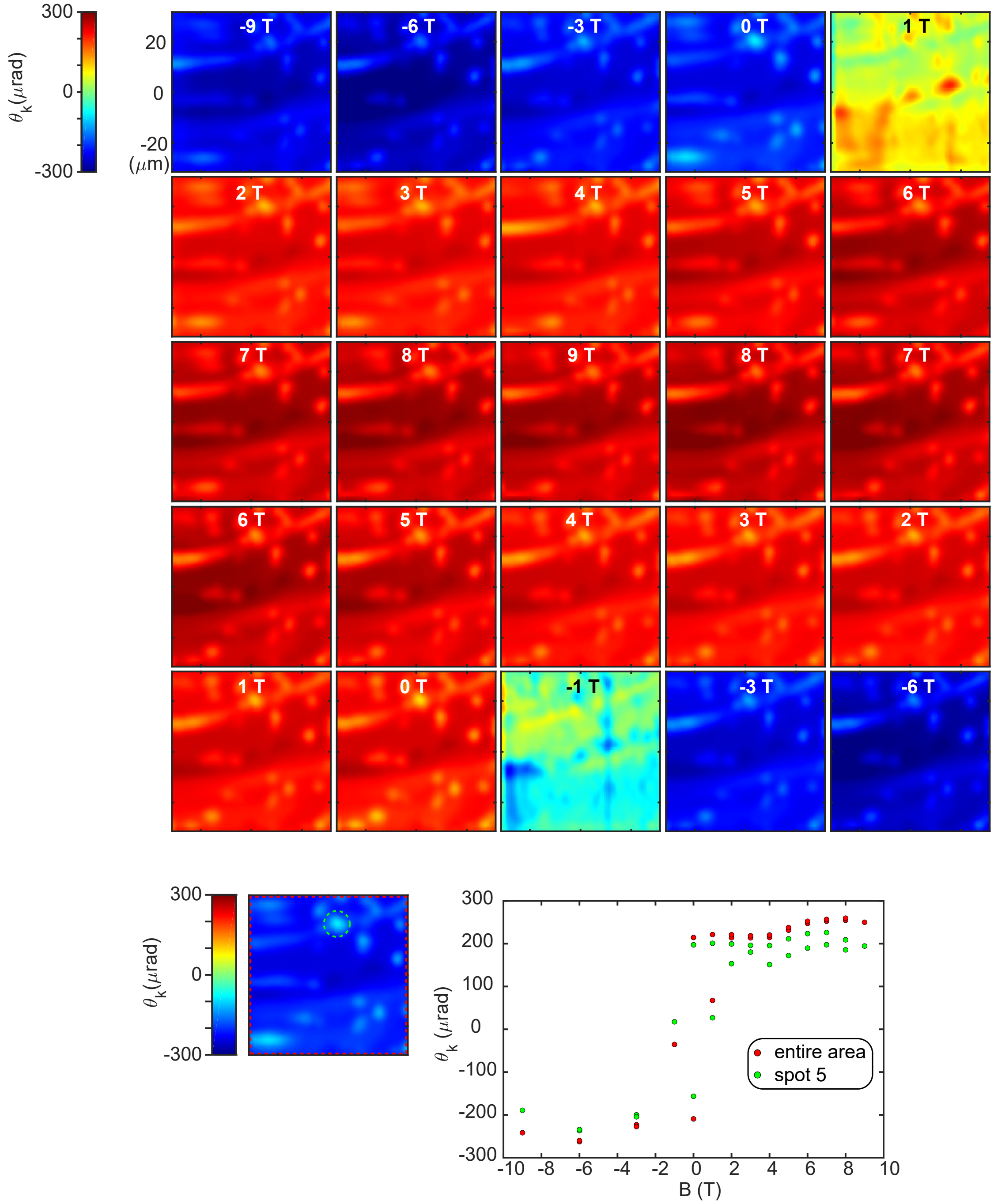

**Supplementary Figure 2. Extended data: Chirality domain switching in another region at 10$K$.** Top: MOKE images taken during hysteresis at $B = 1T$ interval. Bottom: extracted hysteresis loop from the whole area average and from the circled spot 5.

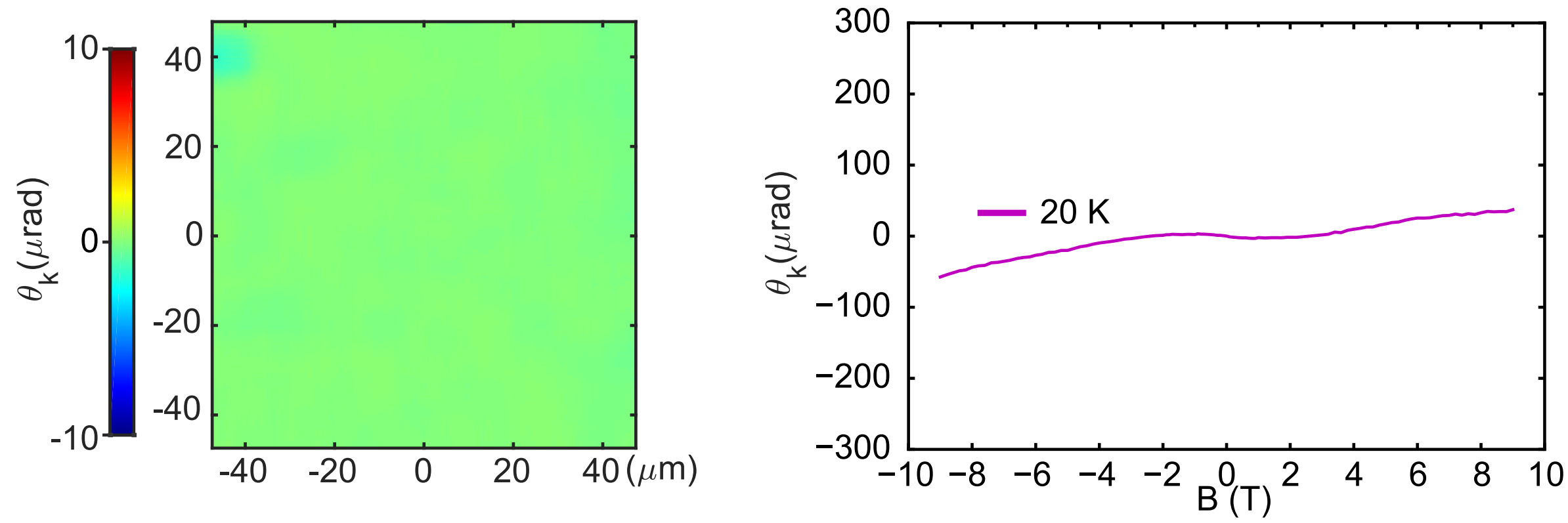

**Supplementary Figure 3. Extended data: Zero-field MOKE imaging and hysteresis at $20\ K$ in the single-Q phase.** No spontaneous $\theta_K$ is present in either MOKE image or single-point hysteresis.

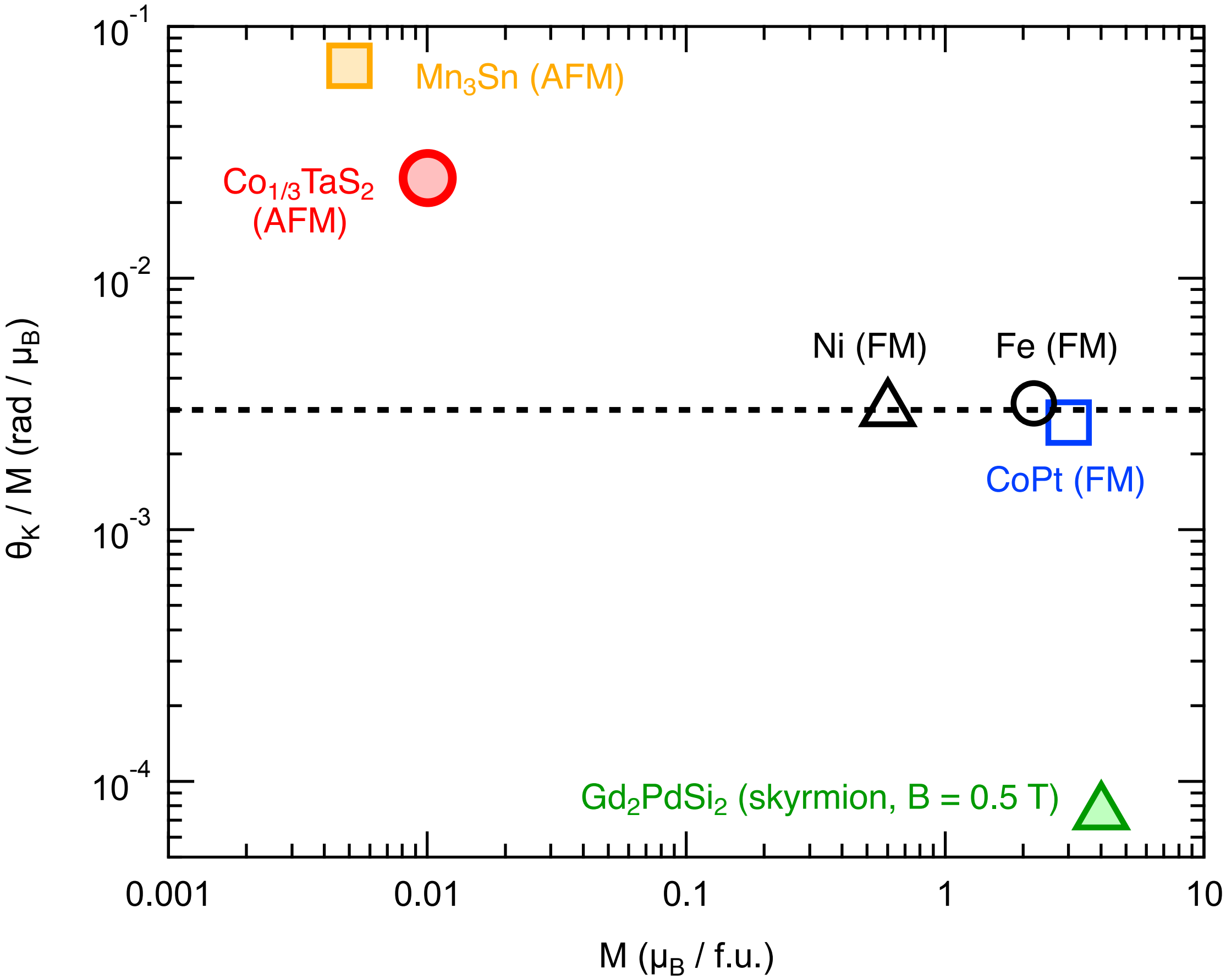

**Supplementary Figure 4. Extended data: MOKE over Magnetization ratio ($\theta_K/M$) of representative ferromagnets (FM), antiferromagnets (AFM), and skyrmion lattice.** Ferromagnets (Ni [1], Fe [1], and CoPt [2]) have spontaneous $\theta_K$ roughly proportional to the magnetization, $\sim 0.03\, rad/\mu_B$ . Noncoplanar antiferromagnet $Co_{1/3}TaS_2$ and coplanar antiferromagnet $Mn_3Sn$ [3] have negligible net moments and are significantly above this ratio. Skyrmion lattice $Gd_2PdSi_2$ [4] is established only within a narrow magnetic field range near $B = 0.5\, T$ and is way below this ratio.

**Sagnac interferometer for MOKE measurements**

The schematics of the zero-loop Sagnac interferometer [5] used in this work is shown in Supplementary Fig. 5. The beam of light from a CW light source centered at 1550 $nm$ is routed by a fiber-circulator to a fiber-polarizer, which polarizes the beam.

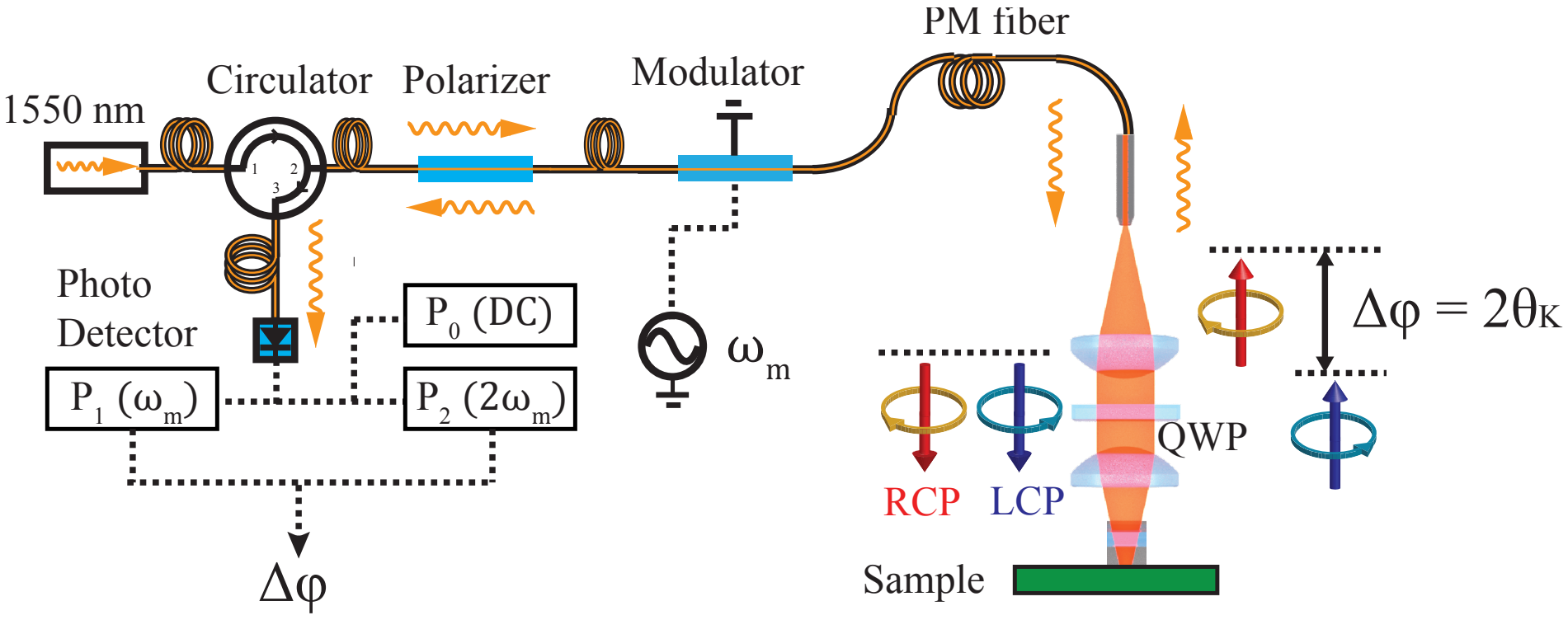

**Supplementary Figure 5. Sagnac MOKE setups operating with continuous-wave (CW) light at 1550 nm wavelength:** Schematics of a zero-area-loop fiber-optic interferometer that is only sensitive to TRSB (MOKE $\theta_K$) effects, which is independent of α. The fiber-optic head can be scanned to simultaneously acquire reflection and MOKE images.

The circulator transmits light from port 1 to port 2 and from port 2 to port 3 with better than 30 dB isolation in the reverse directions. After the polarizer the polarization of the beam is at 45° to the axis of a fiber-coupled electro-optic modulator (EOM), which generates 4.6 MHz time-varying phase shifts $\phi_m$ sin $(\omega t)$, where the amplitude $\phi_m = 0.92$ rad between the two orthogonal polarizations that are then launched into the fast and slow axes of a polarization maintaining (PM) single mode fiber. Upon exiting the fiber, the two orthogonally polarized linearly polarized beams are converted into right- and left-circularly polarizations by a quarter-wave plate (QWP) and are then focused through the optical window of the cryostat onto the sample. After reflection from the sample and passing through the optical window, the same quarter-wave plate converts the reflected beams back into linear polarization with exchanged polarization axes. The two beams then pass through the PM fiber and EOM but with exchanged polarization modes in the fiber and the EOM. At this point, the two beams have gone through the same path but in opposite directions, except for a phase difference of $\Delta\varphi$ from reflection off the magnetic sample and another time-varying phase difference by the modulation of EOM. This nonreciprocal phase shift $\Delta\varphi$ between the two counterpropagating circularly polarized beams upon reflection from the sample is twice the Kerr rotation $\Delta\varphi = 2\theta_K$. The two beams are once again combined at the detector and interfere to produce an optical signal $P(t)$:

$$P(t) = \frac{1}{2} P[1 + \cos(\Delta\varphi + \phi_m \sin(\omega t))] \quad (1)$$

, where P is the returned power if the modulation by the EOM is turned off. For MOKE signals that are slower that the 4.590 MHz modulation frequency used in this experiment, we can treat $\Delta\varphi$ as a slowly time-varying quantity. And $P(t)$ can be further expanded into Fourier series with the first few orders listed below:

$$\begin{aligned} P(t)/P = & \frac{1}{2}[1 + \mathrm{J}_0(2\phi_m)] \\ & +(\sin(\Delta\varphi) J_1(2\phi_m)) \sin(\omega t) \\ & +(\cos(\Delta\varphi) J_2(2\phi_m)) \cos(2\omega t) \\ & +2\,\mathrm{J}_3(2\phi_m)) \sin(3\omega t) \\ & +\cdots \end{aligned} \quad (2)$$

, where $J_1(2\phi_m)$ and $J_2(2\phi_m)$ are Bessel J-functions. Lock-in detection was used to measure the first three Fourier components: the average (DC) power (P0), the first harmonics (P1), and the second harmonics (P2). And the Kerr rotation can then be extracted using the following formula:

$$\theta_K = \frac{1}{2}\Delta\varphi = \frac{1}{2}\tan^{-1}\left[\frac{\mathrm{J}_2(2\phi_m)P_1}{\mathrm{J}_1(2\phi_m)P_2}\right] \quad (3)$$

The noise in Kerr signal is shot-noise-limited to $10^{-7} rad/\sqrt{Hz}$ with $10\ \mu W$ of optical power, which is small enough not to heat up the sample even at the base temperature of the cryostat. By averaging over 100 seconds, 10 nanoradian (nrad) Kerr resolution can be achieved over a few Kelvins variation of sample temperatures. In practice, the bias offset in our system drifts about 20 nrad in experiments that take a long time or over wide sample temperature ranges. And the flexible fiber head can be mechanically scanned to simultaneously produce reflection (P0) and MOKE ($\theta_K$) images.

**Sagnac interferometer's 55 dB rejection of optical anisotropy**

An important feature of the Sagnac interferometer for this work is its exclusive detection of microscopic time-reversal symmetry breaking (TRSB) while rejecting non-TRSB effects such as optical birefringence with 55 dB ($3 \times 10^{-6}$) level of rejection. This is particularly important in $Co_{1/3}TaS_2$ where we have recently discovered a nonvolatile nematic order with a birefringent polarization rotation of $\theta_T = 600\ \mu rad\ sin(2\alpha)$, $\alpha$ being the incident polarization angle [6]. With the 55 dB rejection, we expect the nematic order will introduce at most 0.002 $\mu rad$ false signal in the Sagnac measurements, which is below our sensitivity. This selective sensitivity to TRSB effects is achieved by using a single-mode optical fiber as both the source and detector for counterpropagating, time-reversed light beams. According to Onsager's relations, this configuration guarantees zero signal in the absence of TRSB.

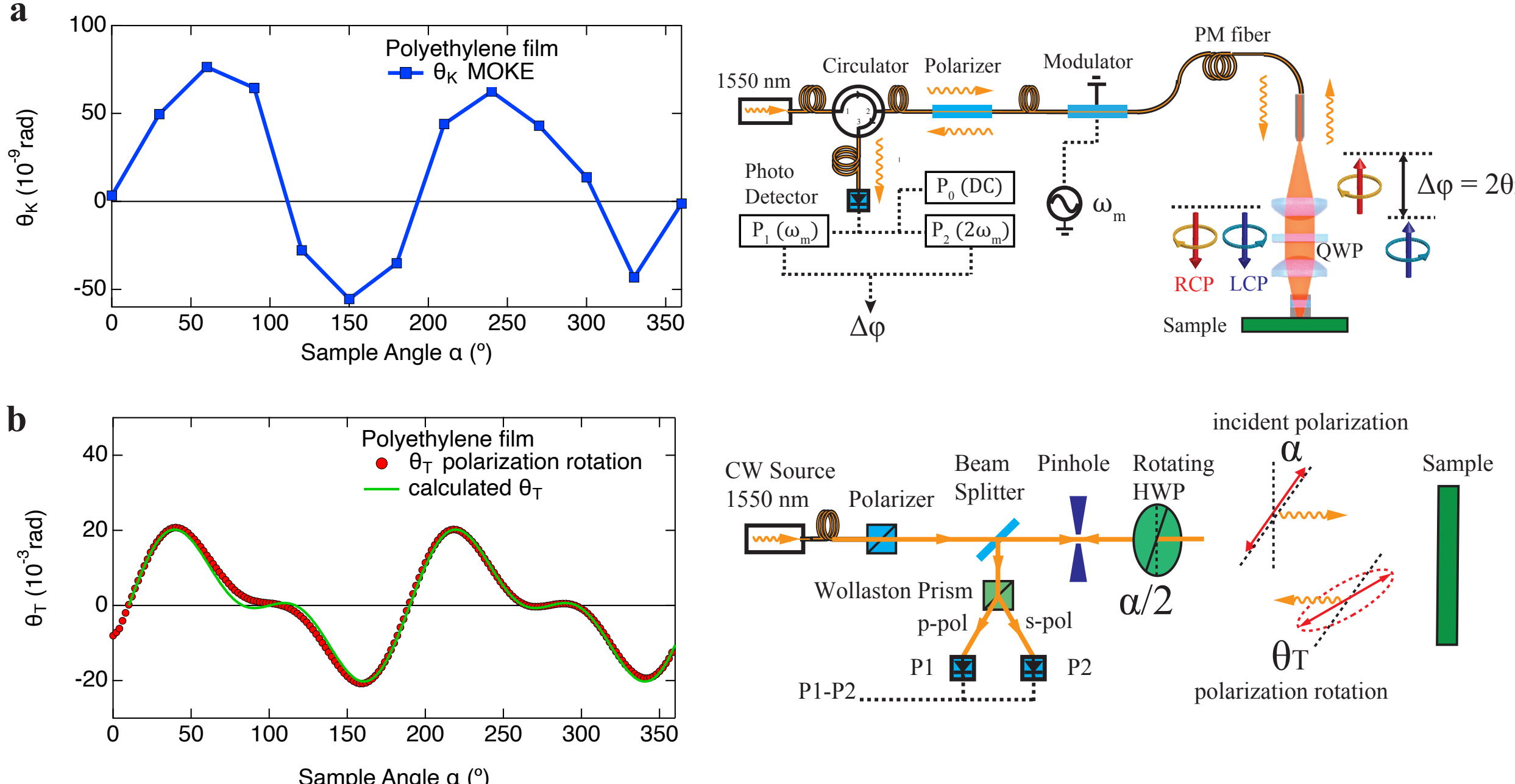

**Supplementary Figure 6. 55 dB rejection of optical anisotropy demonstrated in an anisotropic polyethylene film.** MOKE $\theta_K$ is expected to be zero due to lack of TRSB, while polarization rotation $\theta_T$ is expected in this anisotropy film due to optical linear birefringence (LB) and optical linear dichroism (LD). **(a)** Measured $\theta_K$ by Sagnac interferometry is smaller than 60 nrad. **(b)** Measured $\theta_T$ shows a pattern of 20 mrad that clearly demonstrates an anisotropic (rotational symmetry breaking). Solid line is the calculated polarization rotation for an anisotropic reflective sample, and it agrees well with the measured $\theta_T$. Insets are schematics of Sagnac and polarization rotation measurement setups. Sample angle α is changed by sample rotation with the Sagnac MOKE setup, and by rotating the half-wave plate by $\alpha/2$ in the polarization setup.

The 55 dB rejection of optical anisotropy is demonstrated with a polyethylene film. This polymer film is optically anisotropic and thus produces anisotropic optical rotations. The results are shown in Supplementary Fig. 6. The measured MOKE signal $\theta_K(\alpha)$ (Supplementary Fig. 6a) remains close to zero ($< 60\ nrad$) as the sample doesn't break time-reversal symmetry, while the measured polarization rotation $\theta_T(\alpha)$ (Supplementary Fig. 6b) up to $\pm 20\ mrad$ displays 2-fold rotational symmetry.

The shape of $\theta_T(\alpha)$ in Supplementary Fig. 6b is a direct result of the presence of optical linear birefringence (LB) and optical linear dichroism (LD). And it can be calculated analytically:

$$\theta_T(\alpha) = \ sin(2\,\alpha)cos(2\,\alpha)\ \frac{e^{i\,2\,LB}\,((LD-2)\,LD\,(2+(LD-2)\,LD)\,e^{i\,2\,LB} + (-(LD-1)^2 - e^{i\,4\,LB}\ (LD-1)^2 + e^{i\,2\,LB}(2+(LD-2)\,LD\,(2+(LD-2)\,LD)))}{2+(LD-2)\,LD\,(2+(LD-2)\,LD)+(-2+LD)\,LD\,(2+(\,LD-2)\,LD)\,cos(2\,\alpha))}$$

The calculated $\theta_T(\alpha)$ curve in Supplementary Fig. 6b is obtained using fitting parameters LB = 0.133 and LD = 0.015, which matches well to the experimental data. This polymer film serves as an example that while the polarization rotation setup detects the total polarization rotation, the Sagnac interferometer is sensitive only to TRSB effects.

**MOKE after Zero-field Cool (ZFC)**

In addition to the data presented in Fig. 1d, we performed field-cooling (FC) and zero-field-cooling (ZFC) measurements at a different sample location, as shown in Supplementary Fig. 7. Both the 0.3 T FC (blue) and subsequent zero-field warm (ZFW, red) traces in Supplementary Fig. 7a exhibit a large $\theta_K$ of 180 $\mu rad$, comparable to the 250 $\mu rad$ signal in Fig. 1d from another location. During FC, a slowly varying background arises from the Faraday effect in optical components located inside the magnetic field, as described in Ref.[7]; this background is absent when no external field is applied. The spin chirality domains are aligned by the magnetic field during cooling, resulting in large MOKE signals both in FC and subsequent ZFW.

In contrast, the ZFC measurement (blue in Supplementary Fig. 7b) yields a much smaller $\theta_K$ of $-2\ \mu rad$. Without an applied field during cooling, the chirality domains remain randomly oriented, leading to near-complete cancellation of contributions within the optical probe area.

This contrasting behavior confirms that the MOKE signal originates from the chirality of AFM domains rather than uncompensated moments at AFM domain walls. If the latter were dominant, ZFC that has more AFM domains would produce a larger MOKE signal, contrary to our observations.

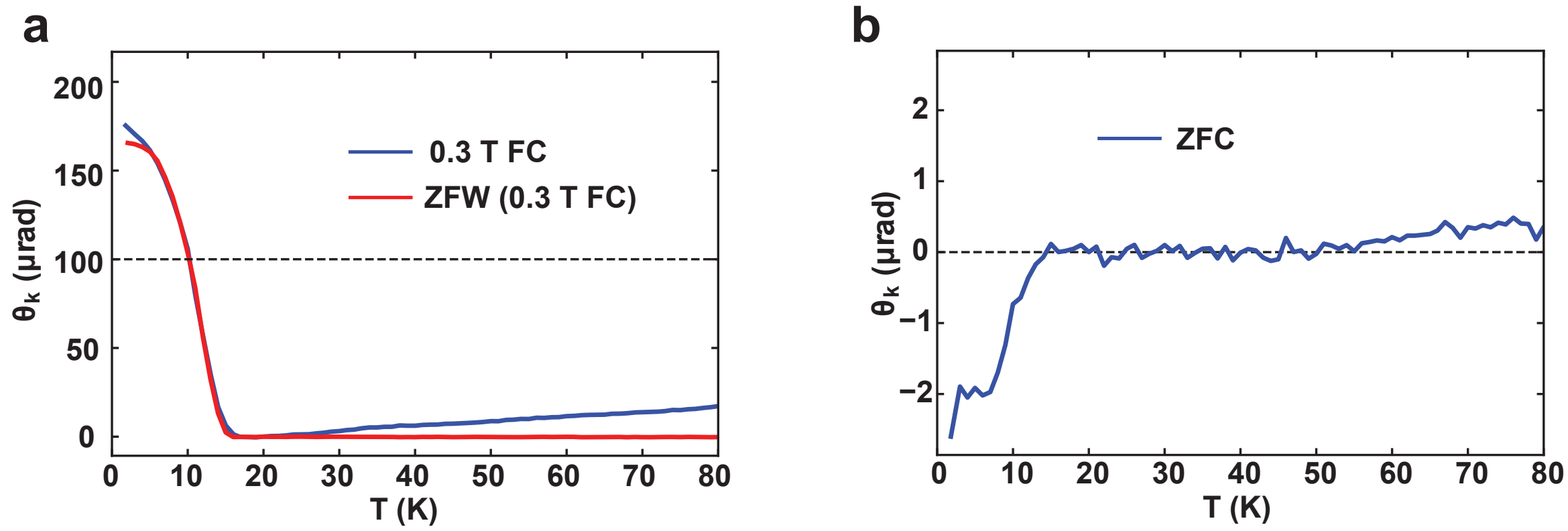

**Supplementary Figure 7. Temperature-dependence of MOKE after field-cool (FC) and zero-field-cool (ZFC) at the same location. (a)** MOKE $\theta_K$ measured during 0.3 T FC (blue) and a subsequent ZFW (red). In the FC (blue), a slowly-temperature-varying contribution is present due to the Faraday effect from optical components inside the magnetic field, which is described in detail in Ref.[7]. Note that such contribution is absent if there is no magnetic field. The spin chirality domains are aligned by the magnetic field during cooling, resulting in large MOKE signal. **(b)** MOKE $\theta_K$ measured during zero-field-cool (ZFC) (blue). The spin chirality domains are randomized, resulting in much smaller MOKE signals due to cancellation from oppositely oriented chirality domains within the optical probe beam.

## Magnetic Force Microscopy (MFM)

We have conducted extensive low-temperature MFM measurements under various experimental conditions. Supplementary Fig. 8 shows a representative scan taken at 5 *K* after zero-field cooling, where AFM domains are expected to be most abundant. However, no magnetic domains or domain walls were detected within the MFM sensitivity. This indicates that the fringing fields produced by uncompensated moments at AFM domain walls are too weak to be resolved by MFM. This is consistent with the picture that the MOKE signal is (dominantly) due to the chirality of AFM domains instead of uncompensated magnetic moments at AFM domain walls.

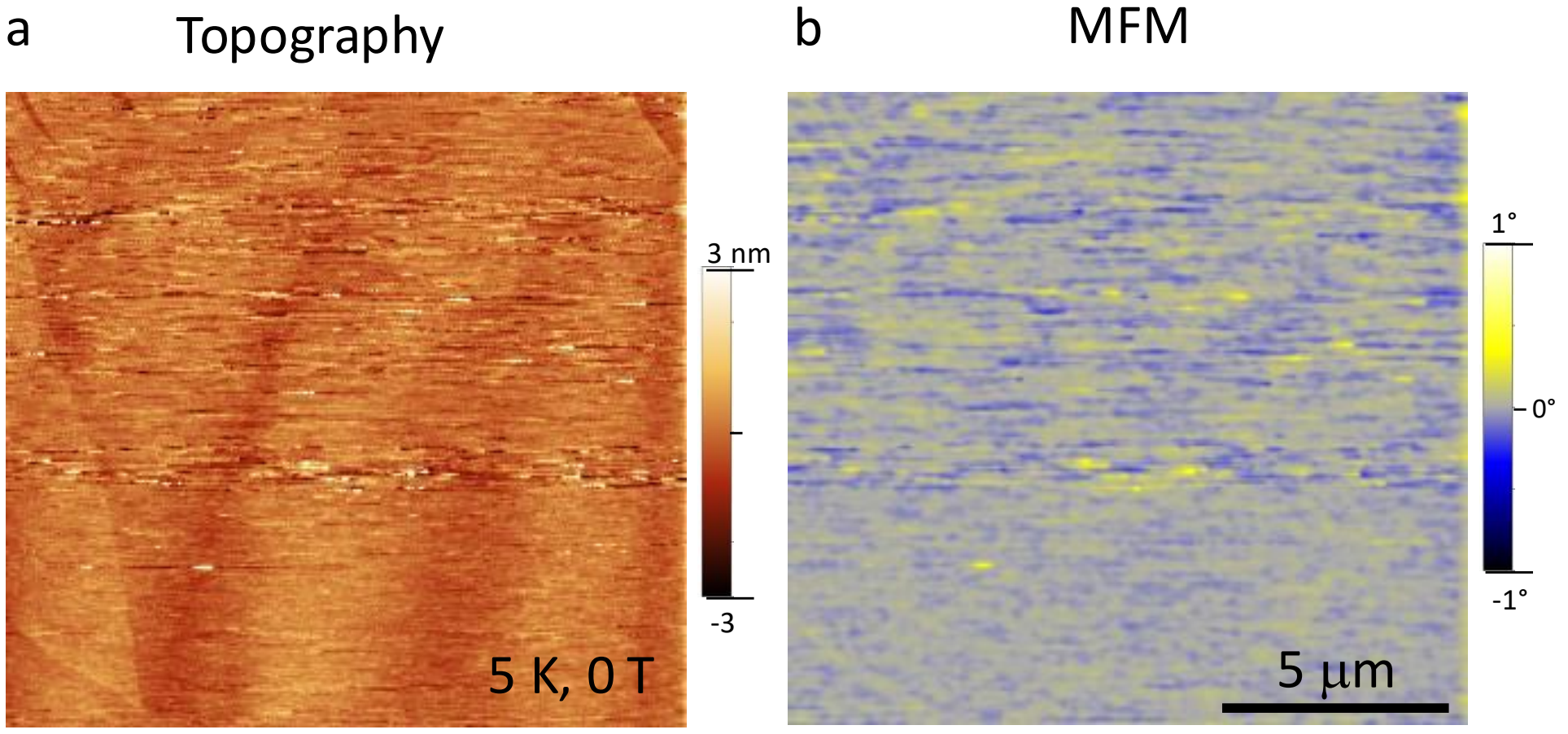

**Supplementary Figure 8. Magnetic Force Microscopy (MFM):** Low-temperature MFM. Topography **(a)** and MFM phase image **(b)** of cleaved $Co_{1/3}TaS_2$ at 5 K after zero-field cooling. No magnetic domains or domain walls are observed within the sensitivity of MFM.

## Energy-dispersive X-ray spectroscopy (EDX)

As discussed in the main text, it is challenging to perform energy-dispersive X-ray spectroscopy (EDX) chemical mapping in the exact same region examined by low-temperature Sagnac microscopy, which provides both MOKE and reflectivity maps at 1550 $nm$. Both techniques are time-consuming, and the millimeter-sized crystal surface is largely topographically featureless, making it impractical to revisit the same tens-of-microns-sized region with different instruments.

Since no known optical transitions exist in $Co_{1/3}TaS_2$ between red light (620 - 750 $nm$) and 1550 $nm$, we assume that the optical reflectivity at 1550 $nm$ qualitatively follows that at red wavelengths. Based on this assumption, we use a wide-field laboratory microscope to measure red-light reflectivity in the same region where EDX was conducted, and infer the corresponding 1550 $nm$ reflectivity.

Supplementary Fig. 9 demonstrates that regions with lower cobalt concentration exhibit higher optical reflectivity at red wavelengths, which we assume also applies at 1550 $nm$. We first locate a flat 500 $\mu m$ -wide region in the electron microscope containing several 40 $\mu m$ pits used as landmarks (Supplementary Fig. 9a). EDX is then performed within four boxes in a smaller 50 $\mu m$ subregion, yielding cobalt concentrations x in descending order: box 10 (x~0.292±0.015) > box 8 (x~0.290±0.015) > box 9 (x~0.289±0.015) > box 7 (x~0.286±0.015), as shown in Supplementary Fig. 9c. The sample is subsequently transferred to the optical microscope, where the same 500 $\mu m$ region is located using the same pits as landmarks, a process that took days (Supplementary Fig. 9b). The red-light reflectivity map reveals very low reflectivity (blue) in the pits and noticeable reflectivity variations (red-yellow-green) even within otherwise flat regions, indicating that the contrast is not purely topographical. Using the pits as reference points, we then image the same 50 $\mu m$ subregion previously used for EDX (Supplementary Fig. 9b). The mean reflectivity R values follow the inverse trend of Co concentration: box 10 (R~0.31) ≈ box 8 (R~0.31) < box 9 (R~0.33) ≈ box 7 (R~0.33). This correlation supports the picture that higher Co content leads to lower red-light reflectivity, an effect which we expect to persist at 1550 $nm$ in the Sagnac measurements.

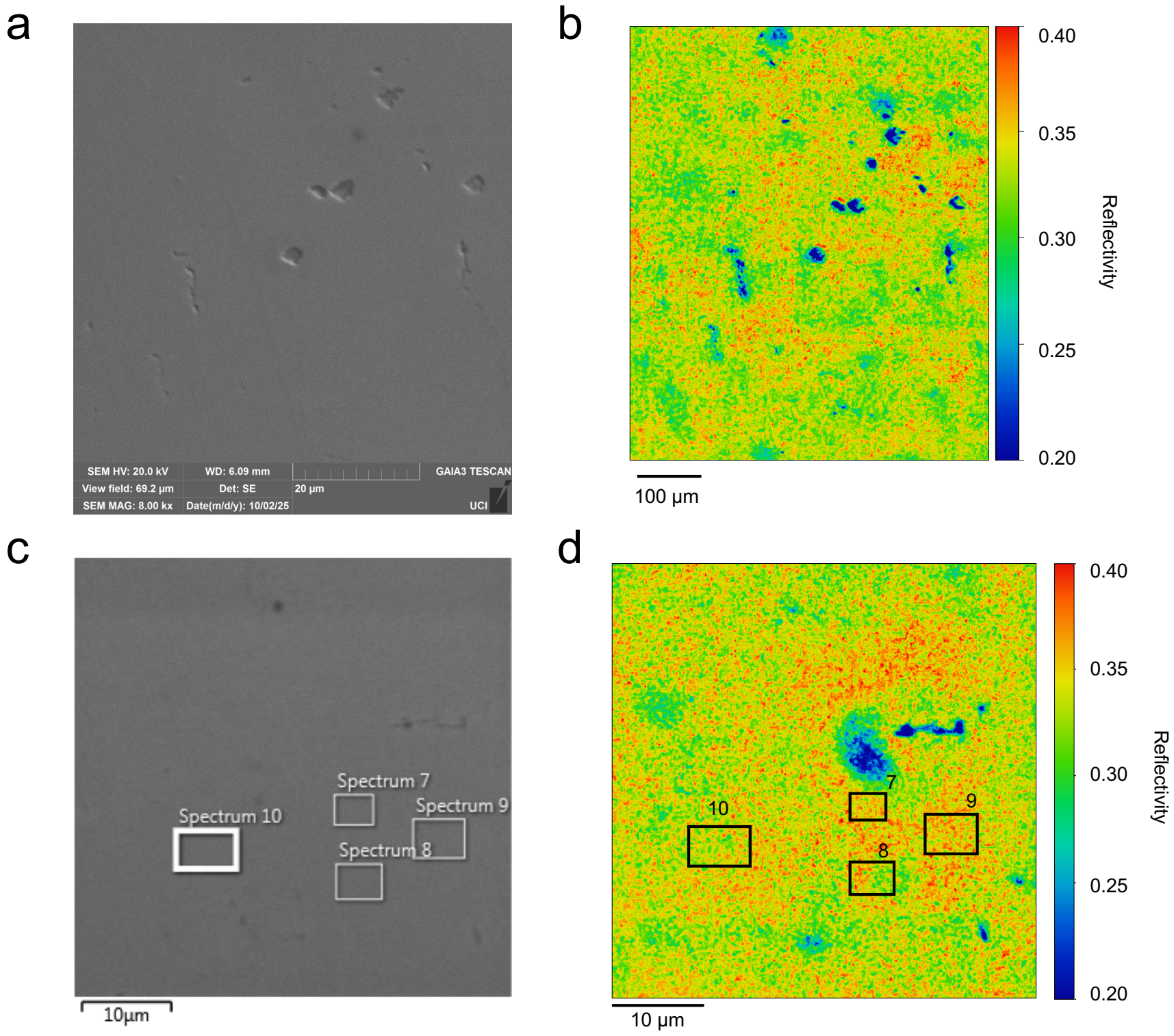

**Supplementary Figure 9. Energy-dispersive X-ray spectroscopy (EDX): (a)** Electron microscope height image in a large region with several 40 $\mu m$ -sized pits as landmarks. **(b)** Red-light reflection image (color coded) of the same region as in **(a)**, located using those pits as landmarks, where reflectivity drops to very low levels (blue). Note that even in flat regions, optical reflectivity varies due to Co-composition. **(c)** Electron microscope image of a small and flat region in **(a)**, with Co-composition determined using EDX inside 4 rectangular boxes: box 10: Co 0.292±0.015 > box 8: Co 0.290±0.015 > box 9: Co 0.289±0.015 > box 7: Co 0.286±0.015. **(d)** Red-light reflection image of the same region in **(c)**, showing lower reflectivity in boxes of higher Co composition: box 10: 0.31 ≈ box 8: 0.31 < box 9: 0.33 ≈ box 7: 0.33.

**Estimating the Contribution of Magnetization to the MOKE Signal**

The conventional mechanism of MOKE ($\theta_K$) is due to the interplay between SOC and band exchange splitting (BES) in the band structure [8] induced by magnetization ($M_Z$) that is either spontaneous (zero-field) or from an external magnetic field. We can estimate this $M_Z$-linear contribution to MOKE in our study, which turns out to be extremely small, less than 1%.

This can be seen by comparing the magnetization $M_Z$ (Fig. 2c) and MOKE signal (Fig. 2d, red) during a magnetic field sweep from 0 to 4 $T$. At zero-field, $M_Z$ (0 T) = 0.01 $\mu_B$, $\theta_K$ (0 T) = 195 $\mu rad$; at 4 Tesla just below the metamagnetic transition, the magnetization increases to $M_Z$ (4 T) = 0.10 $\mu_B$, and MOKE only changes slightly to $\theta_K$ (0 T) = 196 $\mu rad$. This indicated that the MOKE due to magnetization is no bigger than (196 -195) $\mu rad$ / (0.10-0.01 $\mu_B$) = 11 $\mu rad/\mu_B$. Thus, even at 4 Tesla when the magnetization is 0.10 $\mu_B$, the contribution to MOKE from magnetization is at most 11 $\mu rad/\mu_B$ * 0.10 $\mu_B$ = 1.1 $\mu rad$, i.e. only 0.5 % of the total 196 $\mu rad$ signal.

Now consider the simultaneous increase of magnetic susceptibility (Fig. 1b) and zero-field MOKE (Fig. 1d). Both reflect the buildup of spin chirality, but one does not cause the other; they are parallel consequences of the same underlying evolution of spin texture.

A similar argument applies to the field sweep across the metamagnetic transition between 4 T and 6 T (Figs. 2c and 2d). In this region, both magnetization and MOKE increase sharply, by 0.08 $\mu_B$ and 60 $\mu rad$, respectively, due to a sudden reconstruction of the spin configuration (i.e., spin chirality). If the MOKE change were driven primarily by magnetization, it would amount to only 11 $\mu rad/\mu_B$ * 0.08 $\mu_B$ = 0.9 $\mu rad$, i.e., just 1% of the observed 60 $\mu rad$. Therefore, the dominant origin of MOKE is clearly the change in spin chirality, not the net magnetization during the metamagnetic transition.